%% file: ms.tex
\documentclass[10pt,preprint2]{aastex}

\shorttitle{The Ionized ISM in The Central Parsec of the Galaxy}
\shortauthors{Zhao et al.}

\begin{document}

\title{
The High-Density Ionized Gas in the Central Parsec of the Galaxy}

\author{Jun-Hui Zhao\footnote{jzhao@cfa.harvard.edu}, Ray Blundell, James M. Moran}
\affil{Harvard-Smithsonian Center for Astrophysics, 60
Garden Street, MS 78, Cambridge, MA 02138}
\author{Dennis Downes, Karl F. Schuster}
\affil{Institut de Radio Astronomie Millim\'etrique, 38406 Saint Martin d'H\`eres, France}
\author{Daniel P. Marrone\footnote{Hubble Fellow}}
\affil{Department of Astronomy \& Astrophysics, University of Chicago, Chicago, IL 60637}
\email{} 
\email{{\it (Revised: September 2, 2010)}}

\begin{abstract}
We report a study of the H30$\alpha$ line emission at 1.3 mm from the region 
around Sgr~A* made with the Submillimeter Array at a resolution of 2\arcsec\ 
over a field of 60\arcsec\ (2~parsec) and a velocity range of --360 to +345 \kms. 
This field encompasses most of the Galactic center's ``minispiral.'' With an 
isothermal homogeneous HII model, we determined the physical conditions of the 
ionized gas at specific locations in  the Northern and Eastern Arms from 
the H30$\alpha$ line data along with Very Large Array 
data from the H92$\alpha$ line at 3.6~cm and from the radio continuum 
emission at 1.3~cm. The typical electron density and kinetic 
temperature in the minispiral arms are  3-21$\times10^4$~cm$^{-3}$ and 
5,000--13,000~K, respectively.  The H30$\alpha$ and H92$\alpha$ 
line profiles are broadened due to the large velocity shear within 
and along the beam produced by dynamical motions in the strong gravitational 
field near Sgr~A*. We constructed a 3D model of the minispiral 
using the orbital parameters derived under the assumptions that the 
gas flows are in Keplerian motion. The gas in the Eastern Arm appears to 
collide with the Northern Arm flow in the ``Bar" region, which is 
located 0.1--0.2 parsec south of and behind Sgr~A*. Finally, a total 
Lyman continuum flux of $3\times10^{50}$ photons s$^{-1}$ is inferred from 
the assumption that the gas is photoionized and the ionizing photons 
for the high-density gas in the minispiral arms are from external sources, 
which is equivalent to $\sim250$ O9-type zero-age-main-sequence stars.
\end{abstract}


\section{INTRODUCTION}
The Galactic center harbors a supermassive black hole (SMBH)
with a mass of about $4.2\times10^6~M_\odot$ \citep{Ghez08,gill09} 
at the position of the radio source Sgr~A*. The inner parsec of 
the Galactic center region contains a rich cluster of stars, including 
at least 55 OB stars within the central 0.5 pc \citep{paum06}. There are 
several theories to explain how these stars got there. In~situ formation, 
as in giant molecular clouds, is improbable because of the stars' large 
velocity dispersion. The accretion sequence may or may not include the 
formation of an intermediate mass black hole in a super-dense stellar 
cluster that stabilizes the cluster against tidal disruption \citep{hans03}. 
The mystery of the formation of these stars is closely linked to the balance 
of mass inflow and outflow toward the Galactic center and its interaction with 
the source Sgr~A*. Infall of interstellar material may lead to star 
formation up to a certain radius from the SMBH, and the subsequent feedback 
from stellar mass loss may suppress further star formation. Both phenomena 
may determine whether Sgr~A* is in a quiescent state  of the active galactic 
nuclei cycle \citep{loeb04}. 

Earlier radio recombination line (RRL) studies \citep{schw89,rob93,rob96} 
revealed complex structures of ionized gas, including the ``minispiral'' and 
``Bar," whose dynamics remain unclear. Parts of these structures have been 
modeled as gas orbiting Sgr~A* \citep{schw89,sand98,paum04,voll00,lisz03,
muzic07,zhao09} 
while other parts of the structures may be gravitationally unbound \citep{yusef98}. 
Another possibility is that some of the nearby stellar clusters (IRS 13 and 
IRS 16) have powerful winds that, with the help of gravitational focusing, 
create large-scale, unbound flows (e.g., \citealt{lutz93}). There are a number 
of limitations in the existing spectral line observations of these structures. In 
the~cm--radio images made with the Very Large Array (VLA),\footnote{The Very 
Large Array (VLA) is operated by the National Radio Astronomy
Observatory (NRAO). The NRAO is a facility of the National Science Foundation
operated under cooperative agreement by Associated Universities, Inc.} 
the recombination line strength is low, the line-to-continuum ratio is low, 
and the bandwidth is limited, making high-velocity line wings especially 
difficult to detect. In the infrared, the picture given by the 
emission lines from the ionized gas 
is subject to corrections for dust extinction
of $\sim$3 mag in the central parsec \citep{scov03,buch09,scho10}. 
A few key questions about the ionized medium 
associated with Sgr~A* are: \\
\noindent 1) Is the ionized material related to stars or star formation? \\
\noindent 2) Is the ionized material streaming freely in the central 
gravitational potential, or do other mechnisms such as stellar winds and shocks 
influence the dynamics? \\
\noindent 3) What are the column densities (or emission measures), densities, 
kinetic temperatures, and dynamics of the HII gas in the vicinity of Sgr~A* 
itself?

In this paper, we present new observational results on the ionized gas in 
the central parsec of the Galaxy based on observations of the H30$\alpha$ 
line made with the Submillimeter Array (SMA)\footnote{The Submillimeter 
Array is a joint project between the Smithsonian Astrophysical Observatory 
and the Academia Sinica Institute of Astronomy and Astrophysics and is 
funded by the Smithsonian Institution and the Academia Sinica.} and 
we interpret 
them in combination with previous H92$\alpha$ and continuum measurements made with 
the VLA. Section 2 discusses the observations and data reduction. Section 3 
presents our results and the modeling of the physical conditions and 
dynamics of the ionized gas in the vicinity of Sgr~A*. Section 4 
discusses the ionizing properties and possible links to the massive 
star formation in the central parsec, and Section 5 summarizes the 
results and conclusions. We assume a distance of 8~kpc to Sgr~A* throughout 
the paper. At this distance, 1~pc corresponds to $24''$.

\section{OBSERVATIONS AND DATA REDUCTION}

\subsection{{\it SMA H30$\alpha$ Line Data}}

The H30$\alpha$ line observations reported here were obtained with the SMA 
as part of a key project on the Galactic center in the period of 2006-2008. 
Our data comprise eleven 
tracks (Table~1). In all cases, the pointing center was the position of Sgr~A*, 
and the half-power beam width at the 231.901~GHz frequency of the H30$\alpha$ 
recombination line was 53\arcsec\ (2~parsec at the distance of Galactic center). 
The 2~GHz bandwidth of the SMA correlator corresponds to a velocity range 
of $\pm1300$ km~s$^{-1}$. Six tracks were taken in the compact array 
configuration, where the synthesized beam was $5.1''\times3.2''$. 
Five higher-resolution tracks were also acquired,
four in the extended configuration ($1.5''\times1.1''$ resolution) and 
one in the very extended configuration ($0.7''\times0.4''$). 

Calibrations of the data were carried out in Miriad \citep{stw95} with the 
specific implementation for SMA data reduction.\footnote{http://www.cfa.harvard.edu/sma/miriad} The system temperature correction was made in post-processing to 
compensate for amplitude attenuation owing to the Earth's atmosphere. 
Using the calibrators listed in Table~1, we determined the antenna-based 
bandpass solutions for each observing track, and the bandpass corrections 
were applied to the visibility data. The flux density scale was determined 
from observations of planets. Phase corrections were made using the point 
source model of Sgr~A*, conveniently located at the center of the field, 
based on self-calibrations. The continuum emission was subtracted from 
the spectral data using the Miriad task UVLIN for both the point source 
($\sim$3 Jy) and the extended emission ($\sim$3 Jy). 

We constructed the H30$\alpha$ line image cube with robust weighting ($R=2$) 
\citep{dsb95}. The synthesized FWHM beam was $1.9''\times1.5''$ (PA = 26\arcdeg). 
In order to compare the SMA images with the VLA images of the H92$\alpha$ line 
and 22~GHz continuum emission, the SMA image cube was convolved to a circular 
beam of 2\arcsec. The intrinsic resolution of the correlator was 3.2~MHz, or 
4.2~\kms. We smoothed the spectra to a resolution of 15~\kms\ with a Miriad task. 
For this paper, we selected the velocity range of --360 to +345~\kms\ for analysis. 
We found no significant emission beyond this range. The rms noise level of each 
channel image was 7 mJy. The image of the integrated H30$\alpha$ line is shown 
in Figure 1 along with spectra in the directions of known infrared sources. 
Figure 2 shows the distribution of radial velocities determined from fitting 
the peak velocities of the H30$\alpha$ line profile in the central $40''$ of 
the field. Notice the large velocity gradients  along the Northern and 
Eastern Arms. The Western Arc does not show well in the SMA H30$\alpha$ 
data because most of the line emission is below the SMA's 4$\sigma$ sensitivity 
limit. It is detected, however, in single-dish H30$\alpha$
data taken with the IRAM 30m telescope.

\subsection{\it VLA H92$\alpha$ Line Data}
The detailed calibrations and imaging of the H92$\alpha$ line data 
have been  described by \cite{zhao09}. They used the 
line image cube convolved to a circular beam ($\theta_{\rm FWHM}=2$\arcsec),
the same size as that of the H30$\alpha$ line cube. Figure~3 shows a 
comparison of the spectra of the H92$\alpha$ and H30$\alpha$ lines 
toward selected regions containing IR sources. Table 2 lists
the line fluxes and their ratios in these spectra.

\subsection{{\it Continuum Data at 22~GHz}}
The continuum image at 22~GHz was obtained with the VLA high-resolution 
(A and B arrays) data set of 2005 described in \cite{zhao09} 
and the VLA archival  data sets observed in C and D arrays during 
2004 and 2005. The  data reduction procedure described in 
\cite{zhao09} was followed. Sgr~A* and the X-ray transient 
source (J174540.0290031) 3\arcsec~south of Sgr~A* were removed 
from the visibility data prior to imaging. The image of continuum 
emission was constructed from the visibility data with a long-baseline cutoff 
($\le100~{\rm k}\lambda$) to achieve a synthesized FWHM beam of 
$1.95''\times1.41''$. The dirty image was 
cleaned with the Clark--Steer hybrid algorithm and convolved to 
a FWHM beam of $2''$, for comparison with other data at the same resolution. 

\subsection{{\it IRAM H30$\alpha$ Line Data}} 
We also used additional H30$\alpha$ data taken 
with the HERA multibeam receiver on the IRAM 30m telescope, with 11\arcsec\, 
beams, in order to calculate the global parameters for the region. 
Based on the intensity of integrated H30$\alpha$ line emission in 
a lower-angular-resolution ($\sim4''$) 
image constructed by combining the SMA data with the IRAM 30m data, 
we separated Sgr~A West into two components: A, the bright minispiral 
features of the Northern and Eastern Arms ($\int S_{{\rm H}30\alpha}dV\ge5$ 
Jy beam$^{-1}$ \kms) that trace the high-density ionized gas, and  
B, the residuals of Sgr~A West after subtracting the emission from A. 
These results are in Table~2.
\section{RESULTS}

\subsection{{\it Physical Conditions of the Ionized Gas in the Central Parsec}}

The overall radio properties of the observed region (Sgr~A West; $80''\times45''$)
determined from the H30$\alpha$, H92$\alpha$, and 22-GHz continuum images
are summarized in Table 2.
        
\subsubsection{Distribution of H30$\alpha$ Line Emission and IR Sources}

Figures 1 and 2 show the H30$\alpha$ line data in 
the central parsec. Figure~3 shows a comparison between the SMA 
H30$\alpha$ and the VLA H92$\alpha$ line spectra in the direction 
of selected IR sources. For optically thin gas in local thermodynamic 
equilibrium (LTE) with no pressure 
broadening, the ratio of line peak intensity is $\displaystyle 
{R^{\rm H30\alpha}_{\rm H92\alpha}\equiv 
{S_{\rm H30\alpha}/ S_{\rm H92\alpha}} \approx 
{\nu_{\rm H30\alpha}/\nu_{\rm H92\alpha}}} \approx 28$. 
A detailed comparison of the H30$\alpha$ 
line spectra from the SMA and the H92$\alpha$ line spectra observed 
with the VLA yields line ratios in most 
of the regions that are close to the expected LTE ratio, 
suggesting that the ionized gas in the central parsec is 
optically thin and under LTE conditions. 
In some regions, however (e.g., IRS 1W, 10W, 16 and  33), the observed 
ratios show significant departure from the expected LTE ratio.

\subsubsection{Distribution of T$_e^*$}

The equivalent  electron temperature, $T_{e}^*$, derived from the ratio 
of line-to-continuum emission under the assumption that the optically 
thin gas is in LTE, can be estimated from our measurements. In order 
to avoid contamination from both the dust emission, which becomes 
dominant at higher frequencies, and significant synchrotron emission 
at lower frequencies, we used the continuum data at 22~GHz, where 
the contributions from both the synchrotron and thermal dust emission 
are negligible relative to the free-free emission. Figures 4a 
and 4b show the integrated H30$\alpha$ line image and radio continuum 
image at 22~GHz, respectively. We assume that the continuum emission 
at 22~GHz is entirely due to optically thin free-free emission and 
scale its intensity to 231.9~GHz, according to the power law $\nu^{-0.1}$, 
i.e., a factor of 0.79. Using Equation (1) below \citep{wils09}, 
we determined the value of $T_e^*$, 

\begin{eqnarray}
{T_e^*\over K} &=&\Bigg[ {6985 \over a(\nu, T_e)} 
               \left(\nu\over {\rm~GHz}\right)^{1.1}  
               {1\over {1+N(He)/N(H)}}  \nonumber \\ 
& &\left(S_{C}\over {S_{L} \Delta V_{{\rm FWHM}}} \right) \Bigg]^{0.87}~~,
\end{eqnarray}
where $\nu$ is the frequency; $S_L$ is the line flux density; 
$S_C$ is the continuum flux density; 
$\Delta V_{\rm FWHM}$ is the line width (FWHM), and 
only Doppler broadening is involved; 
$N(He)/N(H)$ is the number density ratio, 
which we take to be 0.08; and 
$\alpha(\nu, T_e)\sim0.97$ (at $\nu=22$~GHz and $T_e$, 
the electron temperature, $\sim10^4$~K) 
is the correction to the \cite{mezger67} power-law approximation. 
The angular distribution of $T^*_e$ is shown in Figure~4c. 
$T^*_e$ varies from 7,000 K to 10,000 K in most regions of Sgr~A 
West. However, in the Bar region (e.g., IRS~2L, IRS~12N, IRS~13E, IRS~33),
3\arcsec~SE of Sgr~A*, $T^*_e$ 
increases to $\sim15,000$~K. The significant 
increase of $T^*_e$ in the Bar has also been derived from line observations of
H92$\alpha$ \citep{rob93} and H66$\alpha$ \citep{schw89}.
  
\subsubsection{Isothermal Homogeneous HII Model}      
   
Since the observed quantities of line and continuum intensities depend 
on the density, temperature, and path length of HII gas, we can construct 
a model and fit the data to it over a wide frequency range in 
order to determine the parameters of the ionized gas. In particular, 
with the high-resolution data obtained from interferometer observations, 
uncertainties due to variations in structure and the physical parameters 
across the source are substaintially reduced. Thus, the results from
such an analysis become more reliable. Models for RRL emission from 
the nuclear region of external galaxies have been discussed by 
\cite{pux91,ana93,ana00,zhao96,zhao97,pho98}; and \cite{rodr04}. 
They found that the main constraints for the models are the integrated 
RRL strength at multiple frequencies and the observed radio continuum 
spectrum. The observed quantities (line and continuum intensities) 
depend in a nonlinear way on the distribution of $T_e$ and $n_e$,
the electron density, both along the line of sight and across the 
telescope beam. In the cases of the extragalactic nuclear regions, at 
a given frequency, the line emission often arises from HII components 
under the physical conditions that are particularly favorable at this 
particular frequency. Usually, multiple collections of HII regions 
with different physical parameters are required to fit the observations. 

Both the SMA and VLA data achieve angular resolution (2\arcsec, $\sim0.1$~pc)
adequate to resolve the $\sim1$~pc minispiral structure. 
The effects due to changes of physical 
parameters across the source are therefore less significant. 
However, the electron temperature and density 
could possibly change along the line of sight; $T_e$ and $n_e$ derived 
from the simple isothermal and homogeneous model correspond, 
therefore, to average physical properties of the HII gas in the 
telescope beam. If the HII gas in the beam along the path length 
is isothermal and homogeneous, then the line and continuum flux densities 
from a region subtending a solid angle $\Omega$ are given by (e.g., \cite{shaver75}),
\begin{eqnarray}
S_L&=&{2k\nu^2\over c^2} \Omega T_e 
\Bigg[\left(\tau_L/\beta_n +\tau_C \over \tau_L + 
\tau_C \right) \nonumber \\
          & &\left(1-e^{-(\tau_L+\tau_C)}\right) 
          -(1- e^{-\tau_C})\Bigg]~~,
\end{eqnarray} 
and 
\begin{equation}
S_C = {2k\nu^2\over c^2} \Omega T_e(1- e^{-\tau_C})~~,
\end{equation}

\noindent where $k$ is Boltzmann's constant, $c$ is the speed of light, and 
the line ($\tau_L$) and continuum ($\tau_C$) optical depths are given by

\begin{eqnarray}
\displaystyle
\tau_L &\approx& 575 b_n\beta_n \left(\nu\over {\rm~GHz}\right)^{-1}
\left(n_e\over {\rm~cm}^{-3}\right)^2  \nonumber \\
       & &\left(Lf_V \over {\rm pc}\right)\left(T_e\over {\rm K}\right)^{-5/2}
\left(\Delta V_{{\rm D}}\over {\rm km~s}^{-1}\right)^{-1}  \nonumber \\
       & &\left(1+1.48{{\Delta V_P}\over{\Delta V_D}}\right)^{-1}~~,
\end{eqnarray}

\noindent and 

\begin{eqnarray}
\displaystyle
\tau_C &\approx& 0.08235 \left(n_e\over {\rm~cm}^{-3}\right)^2 
\left(Lf_V\over {\rm pc}\right)  
\left(\nu\over {\rm~GHz}\right)^{-2.1} \nonumber \\
       & & \left(T_e\over {\rm K}\right)^{-1.35} a(\nu,T_e)~~,
\end{eqnarray}

\noindent respectively. $b_n$ and $\beta_n$ are the population departure coefficients; 
$n_e$ is the electron density; $\Delta V_{\rm D}$ and $\Delta V_{\rm P}$ are the
FWHM Doppler and pressure line widths in \kms, respectively;
$L$ is the path length; and $f_V$ is the volume filling factor. 
Note that the last factor in Equation (4) provides a convenient 
approximation for the transition between the pure Doppler 
and pure pressure-broadened cases.

The Doppler broadening has a thermal 
and turbulent component and is given by
\begin{equation}
\displaystyle
\Delta V_{\rm D} = \sqrt{{8{\rm ln(2)}kT_e\over M_{\rm H}}+ (\Delta V_t)^2}~~,
\end{equation}
\noindent where $M_{\rm H}$ is the mass of hydrogen, 
and $\Delta V_t$ is the turbulent width, which here includes 
microscopic turbulence and 
dynamical velocity gradients.
In the central parsec, the thermal width 
($\Delta V_{\rm th}=\sqrt{{8{\rm ln(2)}kT_e\over M_{\rm H}}}$) 
is less than the turbulent width ($\Delta V_t$).
The line width 
due to pressure broadening, computed by \cite{brock71}, is given by
\begin{equation}
\displaystyle
\Delta V_{\rm P} = 1.7\times10^{-18}  
{n_e\over T_e^{0.1}} N^{7.4}~{\rm km~s^{-1}}~~,
\end{equation}
where $N$ is the principal quantum number of the transition. The total 
line width is approximately the quadratic sum of 
$\Delta V_{\rm D}$ and $\Delta V_{\rm P}$,

\begin{equation}
\Delta V_{\rm FWHM} \cong \sqrt{(\Delta V_{\rm D})^2 + (\Delta V_{\rm P})^2}~~.
\end{equation}
$\Delta V_{\rm FWHM}$ is an observable quantity. We solved for the parameters 
$T_e$, $n_e$, $\Delta V_t$, and $Lf_V$ for each point in the image in 
the following way. 
Since $\Delta V_{\rm P}\ll\Delta V_{\rm th}<\Delta V_t$ for
reasonable values of $T_e$ and $n_e$, we assume initial values 
of $T_e$ and $n_e$, and solve for $\Delta V_t$ via Equations (6)--(8). 
We computed the values for $b_n$ and $\beta_n$ for the assumed values 
of $T_e$ and $n_e$ using the NLTE code of \cite{gordon02}, which 
followed the analysis of \cite{walmsley90}. We then formed the 
flux density ratios
$\displaystyle {R^{\rm H30\alpha}_{\rm 22GHz}\equiv {S_{\rm H30\alpha}/ S_{\rm 22GHz}}}$, 
$\displaystyle {R^{\rm H92\alpha}_{\rm 22GHz}\equiv {S_{\rm H92\alpha}/ S_{\rm 22GHz}}}$, and
$\displaystyle {R^{\rm H30\alpha}_{\rm 92\alpha}\equiv {S_{\rm H30\alpha}/ S_{\rm H92\alpha}}}$. 
The use of these ratios eliminates the solid angle $\Omega$ in 
Equations (2) and (3). At each point in the image, we used these 
ratios to solve for $T_e$, $n_e$, and $Lf_V$. We then iterated 
the process, i.e., solved for a new estimate of $\Delta V_t$ with 
these values of $T_e$ and $n_e$. The process converged in a 
few iterations.

The parameters derived for the line of sight for each of the regions are 
summarized in Table 3. Instead of $Lf_V$, we report the emission measure 
($EM= n_e^2Lf_V$). The ratio $T_e/T_e^*$ is given in column 6, which mainly 
corrects for the effect due to underpopulated electrons at the level $N=30$ 
as compared to that expected under the LTE; the ratio 
$\Delta V_t / \Delta V_{\rm FWHM}$ in column 7; and  
$\Delta V_{\rm P}^{\rm H92\alpha} / \Delta V_{\rm D}^{\rm TH}$,
the ratio of the H92$\alpha$ line width of pressure broadening to that of
thermal Doppler broadening, in column 8. The pressure broadening term
($\Delta V_{\rm P}^{\rm H30\alpha}$) is negligible. 

The uncertainties in the physical parameters  mainly come from 
the uncertainties in the measurements. Although the relation between 
the physical parameters and observable quantities is nonlinear, 
we assess the errors through linear approximation. We use the 
optically thin assumption, which appears to be valid at the frequencies 
used in this paper. The fractional uncertainty in
$T_e$, which is proportional to the fractional uncertainties of 
the measured ratios, $\displaystyle 
{R^{\rm H30\alpha}_{\rm 22GHz}}$ and $\displaystyle R^{\rm H30\alpha}_{\rm H92\alpha}$, 
lies in the range from 7\% for IRS 1W to 39\% for IRS 12N. The fractional
uncertainty in $n_e$ is dominated by those of observed line ratios 
$\displaystyle R^{\rm H30\alpha}_{\rm H92\alpha}$ and ranges from 6\% for 
IRS~1W to 36\% for IRS~12N. 

The variations of the H30$\alpha$ line intensities reflect the changes
of excitation conditions in the region. Our model fitting suggests that,
given similar solutions in $n_e$ and $T_e$, 
the large difference in the H30$\alpha$ line intensity
between the two extreme regions (IRS 1W and IRS 7) is mainly due to 
the difference of the emission measure (EM) or the path length along 
the line of sight (see Table 3).

In short, the line ratio of H30$\alpha$ to H92$\alpha$ varies from
15$\pm$4  to 32$\pm$4. The derived mean values of the 
line optical depth are $\overline{\tau}_L{\rm (H30\alpha)} = -0.001$ and 
$\overline{\tau}_L{\rm (H92\alpha)} = -0.006$ with standard deviations
of $\sigma_{\tau_L{\rm (H30\alpha)}}=0.0004$ and 
$\sigma_{\tau_L{\rm (H92\alpha)}}=0.005$ from the corresponding emission
regions. The free-free optical depth at the H30$\alpha$ line 
frequency $\tau_C{\rm (231.9~GHz)}$ varies from 1$\times10^{-4}$
to 1$\times10^{-5}$  while 
$\tau_C{\rm (8.3~GHz)}$ is in the range between 0.16 and 0.01. Because of the small optical depths
in both line and continuum, the non LTE effects in the minispiral, in general,  are
not  critical.

There is a diffuse nonthermal synchrotron component in the central parsec 
region \citep{ekers83}. Based on the flux density of 22 Jy at 20~cm in 
the central $40''\times70''$ from \citet{ekers83}, we estimated 
that the contribution to the continuum flux density in a 2\arcsec~beam 
is $\sim$10 mJy under the assumption that the nonthermal spectral index 
$\alpha\sim-0.5$ ($S_\nu\propto\nu^\alpha$) and the emission are evenly 
distributed. The optical depth of the H92$\alpha$ line is small 
($\tau_L\sim -0.01$). The contribution from the stimulated H92$\alpha$ 
emission of $\sim0.1$ mJy is insignificant in the regions discussed 
in this paper. The contribution from the stimulated 
line emission by the dust emission at 232~GHz is negligible because 
of the small optical depth of the H$30\alpha$ line.

The observed line widths of the HII gas in the minispiral arms at the 
Galactic center are dominated by $\Delta V_t$, which is 
consistent with velocity gradients due to gas motion around Sgr~A* or the 
SMBH being primarily responsible for the broadening in the line
profiles observed in the 2\arcsec\ beam.

\subsubsection{Northern and Eastern Arms}
The ionized gas of the high-density components in the Northern and Eastern 
Arms has a mean kinetic temperature of $\overline T_e\approx 6\times10^3$ 
K and an electron density of $ n_e$ in the range of $10^{4-5}$~cm$^{-3}$, 
which is consistent with the values ($10^4$~cm$^{-3}$) derived from 
Pa$\alpha$ line emission at 1.87 $\mu$m \citep{scov03} for a given  
uncertainty in extinction corrections in the near IR. The HII gas properties in 
the Northern and Eastern Arms are similar to these in HII regions 
around young and/or evolved massive stars. 

\subsubsection {The Bar}
We confirm that the ionized gas in the Bar 
(e.g., IRS~2L, IRS~12N, IRS~13E, and IRS~33 in Table 3)
appears to have a higher kinetic temperature in the range of 
$8.5\times10^3$K $\le T_e\le1.3\times10^4$K, as suggested by 
earlier RRL observations \citep{rob93}. The electron 
density shows a range from a few times 10$^4$ to a few times 10$^5$~cm$^{-3}$. 

The heating process may be complicated in the Bar because of 
complex gas interactions. There are several energy sources in the 
region. First, strong winds from the central stellar clusters such 
as IRS 16 and IRS 13 could provide kinetic energy to compress the 
ionized gas in this region. The heating due to the interaction 
between the winds and the ionized gas in the Bar likely 
occurs in the regions localized to the cluster sources. 

In addition, 
the well-known minicavity (marked with the gray circle in Figures 
4a and 4b), a region depressed  of radio continuum, is located 
3.5\arcsec~SW of Sgr~A* \citep{yusef89}. 
This region is likely 
created by a fast wind outflow from the sources in the vicinity
sweeping up the gas in the orbiting ionized streams, as suggested 
by the morphology of a bright [FeIII] line in the infrared 
\citep{lutz93}. 
However, the minicavity is a few arcsec displaced 
from the high-kinetic-temperature region, which in fact
is best seen as a depression associated with H30$\alpha$ line emission.
The minicavity can also be explained  by \cite{melia96}
as a downstream, focuced flow due to the accretion by the SMBH at
Sgr A*.

Alternatively, the high temperature in the Bar could result from a 
collision between two ionized orbiting flows, as suggested in the 
dynamical model of \cite{zhao09}. A computation using their 
Keplerian orbital model (see Section 3.2) shows that the specific kinetic energy 
reaches a maximum (see Figure~4d) in the Bar region where high 
kinetic temperatures  (see Figure~4c) are observed. The distribution 
of specific kinetic energy (Figure~4d) shows a substantial increase 
from $2\times10^{14}$ erg~g$^{-1}$ at a region near IRS~5 in the
Northern Arm to $12\times10^{14}$ erg~g$^{-1}$ at IRS~33 in the Bar.
However, we note that a high kinetic energy of the orbiting flow 
does not have to correspond to a high kinetic temperature; the 
ionized flows can release their kinetic energy to shocks if the 
ionized flows collide with each other. If they collide with the 
strong stellar winds from the clusters of stars in the Bar area,
then the gas could be heated by the shocks to  higher temperatures.
In addition, the model also shows that the ionized flows of the 
Northern and Eastern Arms do collide at the region with  high 
kinetic temperatures (see Figure~21d of \cite{zhao09} and the 
3D structure of the minispiral arms in Figure~5). The observed 
high kinetic temperatures in the Bar suggest that the ionized gas 
could be further heated by the shocks generated from the collision. 

\subsection{{\it Kinematics and Dynamics}}

Kinematics of the ionized gas in both the Northern Arm and Western Arc 
have been explained as Keplerian circular motion of a circumnuclear 
ring \citep{schw89,lacy91,voll00,lisz03}. Based on the VLA observations 
of H92$\alpha$, \cite{rob93} found that the ionized ring consists of 
three separated kinematic components, the Western Arc, Northern Arm, 
and Bar. The radial velocities of the Western Arc can be fitted with 
a Keplerian model with circular motion.  Using the observations of 
the Brackett-$\gamma$ line, \cite{paum04} found that the kinematics 
of the ionized gas in the Northern Arm can be explained as a bundle 
of elliptical orbits. In addition, the Eastern Arm may be interrupted 
by a high-velocity component ``tip''. Also, the Bar component 
extends to the IRS6/34 region which is affected by two other kinematic 
features, the ``Western Bridge'' and the ``Bar Overlay.'' The VLA 
observations of the H92$\alpha$ line suggest that the loci of the three 
minispiral arms can be fitted with three families of elliptical orbits; 
the orbits of both the Northern and Eastern Arms have high eccentricities 
while the Western Arc is in nearly circular motion \citep{zhao09}. The 
VLA measurements of the proper motions and the radial velocity resolved the 
degeneracies in the inclination angles, suggesting that the two ionized 
flows from the Northern and Eastern Arms collide in the Bar region.

On the other hand, \cite{muzic07} suggested that the non-Keplerian motions
might be important in the kinematics of the filaments observed near Sgr A*. 
In addition, \cite{voll00} noticed
that a substantial amount of velocity dispersion along with additional inwards radial
motion is required in order to fit the observed velocities of the ionized
flows with their Keplerian circular-orbit model.

\subsubsection{3D View of the Minispiral Structure and the 
Circumnuclear Disk}

Using the Keplerian model described in \cite{zhao09} along with the orbital 
parameters derived from fitting the VLA H92$\alpha$ data and proper motion
measurements, we recomputed the 3D locations of the three ionized arms. 
Figure~5 shows a 3D visualization of the three ionized flows in the Northern,
Eastern, and Western Arms (Arc), which illustrates that the Northern Arm 
and Western Arc are nearly coplanar based on the similar longitudes 
of their ascending nodes ($\Omega$) and inclination angles ($i$) derived 
from fitting the loci of the ionized flows. The mean values of the 
two orbital angles,
$\overline \Omega=71\arcdeg\pm6$\arcdeg~and $\overline i=119\arcdeg\pm3$\arcdeg,
are inferred for the two coplanar orbits (Northern Arm and Western Arc), 
which form the ionized ring. The orbital plane of the ionized flows 
appears to be aligned with the 
plane of the circumnuclear disk (CND) derived by \cite{gust87} from 
molecular line observations. If the fitted parameters from both the 
molecular lines \citep{gust87} and hydrogen  recombination lines 
\citep{schw89,lacy91,rob93,voll00,lisz03,zhao09} are correct, 
the Keplerian motions dominate the kinematics in the 
central parsec, and the plane of the main nuclear gaseous disk or the 
CND can be described by the mean orbital angles inferred above. 
Therefore, the orbital plane of the Eastern Arm 
($i=122\arcdeg\pm5\arcdeg$ and $\Omega=-42\arcdeg\pm11\arcdeg$) is 
nearly perpendicular to that of the CND. Figure~5 shows that the 
Eastern Arm flow runs into the CND from the northeast and the near 
side with respect to Sgr~A* and collides with the Northern Arm 
flow in the region located south and behind Sgr~A*.
  
\subsubsection{Kinematics and Keplerian Motions}

Figure~6 shows the spectra taken from 12 positions along 
each of the two ionized arms (shown in Figure~7). The observed spectra 
are compared with the radial velocities (green bars) derived from the 
Keplerian model. Note that we assume that the LSR velocity of the black hole,
$V_z$, is 0 to within $\pm$ 5 km s$^{-1}$ (see discussion in \cite{gill09}). 
In general, the kinematics of the ionized gas follow the 
Keplerian orbital motion expected for the gravity of the  SMBH 
($4.2\times10^6~M_\odot$) at Sgr~A*. Models of the enclosed mass
(see \citealt{okf09}) show that the added enclosed mass starts to 
rise above the black hole's mass starting at a radius of 0.3 pc (7\arcsec) 
from Sgr A*.  At a radius of 1 pc (24\arcsec) from Sgr A*, near the tip 
of the minispiral seen in projection, the enclosed mass has nearly 
doubled, to about $8\times10^6~M_\odot$. Thus, for a given circular
orbit with a radius of 1 pc the actual orbital velocity would be about 1.4 times 
higher than in a pure Keplerian approximation based on the mass of 
the black hole only.  
Apart from this effect, however, in several regions the radial 
velocities deviate even more significantly
from those predicted from the Keplerian model.

\subsubsection{Non-Keplerian Components}
In the northwest end of the Eastern Arm near the location E17 
(Figure~7a), a feature with radial velocities blue-shifted up to 
$\sim -200$~\kms\ shows a large deviation from the value of 
$\sim-100$~\kms\ predicted by the Keplerian model (Figure~6). 
From \cite{paum06}, we find that at least five massive stars---IRS 34W 
(Ofpe/WN9), IRS 34E (O9--9.5I), IRS 34NW (WN7), IRS 7SW (WN8), 
and IRS 3E (WC5/6)---are located in a region with a radius of 
2.2\arcsec\ (indicated with a large black circle near region
E16 in Figure~7a), referred 
to as IRS 34 hereafter. Except for IRS 3E, which has no proper 
motion measurements, the proper 
motion data of the remaining four stars \citep{paum06} show that 
the stars all move southwest toward the ionized flow with a mean 
transverse velocity of 185 \kms\ (PA = --134\arcdeg) (as shown by an arrow 
in Figure~7a). We compare the radial velocity of the stars 
with the spectrum (inset c in Figure~7a) of the H30$\alpha$ line 
emission integrated from the IRS 34 region. 
The four stars besides IRS 3E show negative radial velocities in the 
range \hbox{--150} \kms\ to \hbox{--340} \kms, which is blue-shifted with respect to 
the peak velocity ($-146\pm4$ \kms) of the broad 
($\Delta V_{\rm FWHM}=160\pm8$ \kms) spectral feature. 
The stellar winds from the Wolf-Rayet (WR) stars, evolved from 
early type O-stars with mass loss rate of 10$^{-5}~M_\odot$ yr$^{-1}$, 
appear to play a considerable role in the local kinematics. The strong 
stellar winds with velocities up to a few times 10$^{3}$ \kms~could sweep 
up the ionized flow of the Eastern Arm, causing the peculiar velocities 
and the velocity gradient across the Eastern Arm at the northwest end.
In fact, the motions of the filaments associated with the minispiral
found by \citep{muzic07} in infrared also suggest that a model
of purely Keplerian motions of the gas
is inadequate and the filaments could be due
to the interaction of a fast wind with the minispiral. 

In the tip region \citep{paum04} with peculiar radial velocity that 
deviates signficantly from the Keplarian 
velocity by +150 \kms\ (see panel E12 in 
Figure~6 and Figures 7a and 7b), there is also a WN8 star, IRS 9W. 
The radial velocity (+140 \kms) of IRS 9W \citep{paum04} is similar 
to that of ionized gas in the Eastern Arm. A comparison between 
the stellar transverse velocity of 215 \kms\ (PA = 51\arcdeg) 
and that of 304 \kms\ (PA = --67\arcdeg) predicted from the 
Keplerian model for the ionized flow at the location of IRS 9W 
suggests that the star does not follow the orbit of the Eastern 
Arm but may run across the Eastern Arm. Then, the high-radial velocity 
tip is likely to be the part of the gas in the ionized flow that is 
compressed and accelerated by the suspected strong wind from the 
IRS 9W star.  The velocity discontinuity at the tip may be evidence
for the presence of shocks in the region. 
  
In addition, deviations from Keplerian motions are also observed in the 
Bar region. The non-Keplerian kinematics could be attributed both to the 
interaction between the ionized flows in the  Northern and Eastern Arms
and to the interaction of the ionized flows with the strong winds from the 
massive star clusters (the IRS 16 and IRS 13 clusters, for example). 
Well-organized polarized emission from magnetically aligned dust grains in the central
parsec has been observed \citep{aitk91,aitk98,glas03} suggesting
a magnetic field strength of $\geq$2 mG. The corresponding energy 
densities in the magnetic fields
are $\sim1.6\times10^{-7}$ erg cm$^{-3}$, about twice larger
than the thermal energy densities in the ionized gas 
($T_e\sim7\times10^{3}$ K and
$n_e\sim6\times10^{4}$ cm$^{-3}$ at IRS 1W) but
two orders of magnitude less than the kinetic energy densities
of the orbital motions. The magnetic fields frozen in the ionized
streamers can not substantially alter the bulk motions of the gas but produce
some cumulative magnetohydrodynamic (MHD) effects, such as
orbital compression of the streamers as implied from the observed
evidence for the convergence of field lines near Sgr A* \citep{glas03},
and a helix of a natural morphology for a 
twisted magnetic field frozen into a plasma as observed 
in the Northern Arm \citep{yusef89,zhao09}.
In the region close to Sgr~A*, the magnetic field probably becomes stronger, 
and the MHD effects may also play a significant role in 
the kinematics of the ionized flows. 

An interesting anomalous feature located $\sim2''$~northwest of 
Sgr~A* can be seen in the integrated H30$\alpha$ line intensity image 
(Figure~1). The radial-velocity image (Figure~2) indicates a 
large velocity gradient across this feature. The H30$\alpha$ line 
spectrum (inset d in Figure~7) suggests that this feature is 
composed of at least three spectral components ($\geq4\sigma$) 
at radial velocities of 160  \kms, 39  \kms, and --250 \kms, with a 
line width of $\sim50$~\kms, a typical velocity width of 
the ionized gas in the minispiral arms. We note that there is a 
compact continuum source called $\varepsilon$ \citep{yusef91}, 
located 1\arcsec~south of this anomalous feature, corresponding 
to a hole in the H30$\alpha$ line images (e.g., Figure~7a and Figure 1). 
The source $\varepsilon$ appears to be moving away from Sgr~A* with 
a transverse velocity of 340 km s$^{-1}$ \citep{zhao09}, which 
has been interpreted as the confluence of gravitationally 
focussed winds from the IRS 16 cluster \citep{ward92}. The 
H30$\alpha$ line property of the anomalous feature appears to 
be consistent with the focused-wind model, suggesting
that the velocity components in the anomalous feature correspond to 
ionized blobs of the minispiral arms that have been 
blown out of the region  $\varepsilon$
due to the high pressure of the confluent stellar winds.
However, given the closeness to Sgr~A*, the detection
of this feature needs to be further confirmed with 
higher-sensitivity observations. 

Finally, in the locations with large angular offsets from Sgr~A*, we noticed that 
the deviations in radial velocity from the Keplerian motions become 
significant for the Northern Arm ($>50$ \kms), as shown in both Figures 6 
and 7, while for the Eastern Arm, the deviations are less significant.
The difference in the deviation of the observed radial velocities
from the Keplerian velocities between the Northern and Eastern Arms
suggests that the distribution of mass has a preference for the plane of the CND.

\section{IONIZATION}
\subsection{\it The High-Density Gas in the Northern and Eastern Arms}
We have shown that  the H30$\alpha$ lines observed with the SMA 
are excellent tracers of the high-density HII components in 
the Northern and Eastern Arms. Based on the observations and derived
physical parameters, we can estimate the ionizing photon rate ($Q=n_e^2\alpha_BV$)
required for maintaining the ionizations in these regions for the 
case where the ionized gas is optically thick in the Lyman lines 
and where $\alpha_B$ is the ionization coefficient, and $V\approx\Omega D^2Lf_V$ 
is the volume of the ionized clumps. The ionizing photon rate can be written as

\begin{eqnarray}
Q(H)  &=& 
     1.31\times10^{45}
              \left[\alpha_B(T_e)\over\alpha_B(10^4~{\rm K})\right] \nonumber \\
            &&\left[\theta_B\over {\rm ~arcsec}\right]^2 
              \left[n_e \over 10^4 {\rm ~cm^{-3}}\right]^2 
             \left[D \over 8 {\rm ~kpc}\right]^2  \nonumber \\ 
            && \left[Lf_V\over 10^{-3} {\rm ~pc}\right]~{\rm phot.~s}^{-1} ~~,
\end{eqnarray}
where $\theta_B$ is the resolution of the observations (beam) with the 
assumption that source fills the beam; and $\alpha_B(T_e)$, the total 
recombination coefficient to excited levels, is given by \cite{humm63}. 
We calculated the values of 
the inferred ionizing photon flux required to maintain the ionization 
of the high-density components in the Northern and Eastern Arms as well 
as the HII mass ($M_{\rm HII}$) and excitation parameter ($U$). On the 
assumption that the HII gas is internally ionized, the equivalent 
zero-age main sequence (ZAMS) star that is the ionizing source was estimated. 
The internal-ionization model requires at least $\sim25$ O9-type stars embedded 
and roughly evenly distributed in both the Northern and Eastern Arms.   

\subsection{Ionizing Sources}

We now examine where the known OB stars lie in the region.
Figure~8 shows the 90 massive stars within a radius of 14\arcsec~(0.5 pc) 
of Sgr~A* identified by \cite{paum06}, among which 55 are OB stars. 
The 14 OB stars within a radius of 0.85\arcsec\ (0.03 pc) from Sgr~A* 
are known as S stars. No obvious correlations appear between 
the location of 41 OB stars outside the radius of 0.85\arcsec~and 
the distribution of the high-density HII gas in the minispiral 
arms on the projected plane. If we assume the average solid angle for the 
minispiral subtended to be $\Omega_s$, then the required number of 
ionizing photons from the stars will be

\begin{eqnarray}
Q^*(H)=Q(H)(4\pi/\Omega_s)~~.
\end{eqnarray}
If we assume $4\pi/\Omega_s\sim10$, then 250 O9-type stars are required.

For the lower-density ($\overline {n}_e\sim 8\times10^2$~cm$^{-3}$) 
component B, the ionized mass of $\sim340~M_\odot$ is 
inferred, requiring  a Lyman photon flux of $1.2\times 10^{50}$
phot. s$^{-1}$ or $\sim$100 O9-type stars (ZAMS). The 250 O9-type stars 
required for externally maintaining the ionization of the minispiral 
appear to be adequate in supplying the leaking  Lyman photons from 
the minispiral to further ionize the lower density component that 
pervasively distributes in the region. Thus, the ionizing source to 
maintain the ionization state of Sgr~A West is equivalent to $\sim$250 
O9-type stars (ZAMS). Considering that a substantial amount of
the ionizing photons can be produced by the hot evolved massive stars
and the survey of \cite{paum06} is 
incomplete, there may be enough stars to ionize the region.

The origin of the ionized minispiral arms and the birthplace of the 
massive stars found in the central parsec are still matters of 
conjecture. There are two possible models that may explain the 
presence of the massive stars near the SMBH. First, the stars 
formed elsewhere outside  the central parsec and migrated to 
the Galactic center \citep{gerh01}. However, compared with the 
age of the massive stars in the central parsec, it takes too long 
for stars to move to the center from several parsec away  by losing 
their orbital energy through dynamical friction \citep{porte03,kim03}. 
Alternatively, the stars may have formed near the SMBH by an unusual 
astrophysical process, such as the fragmentation of a massive disk 
rotating near the SMBH \citep{nayak07}. Recent modeling and numerical 
simulations 
\citep{nayak07,ward08,br08} show that for an infalling giant molecular
cloud (10$^{4-5}~M_\odot$) interacting with an SMBH 
(\hbox{1--$3\times10^6~M_\odot$}),
a small fraction of the gas cloud becomes bounded to 
the black hole, forming an eccentric disk that quickly fragments 
to form stars. The tidal disruption of the self-graviting
clouds formed coherent spiral structures, i.e., the 
minispiral arms, which are ionized by
the Lyman continuum photons produced from the newly formed massive stars. 

\section{SUMMARY AND CONCLUSION}
We report results from the SMA observations of the H30$\alpha$ 
line emission from the Galactic center at a resolution of 2\arcsec. 
We measured the velocity profiles over the central $60''$ (2 parsec) 
region around Sgr~A*. We carried out an analysis of the line and 
continuum data  based on an isothermal homogeneous HII model for each 
component. 
With the constraints from the SMA H30$\alpha$ data at 1.3 mm and 
the VLA H92$\alpha$ data at 3.6~cm and continuum data at 1.3~cm, 
we determined the physical conditions of ionized gas in the 
Northern and Eastern Arms of the minispiral. We found that the 
typical electron density and kinetic temperature in the arms are 
\hbox{3--$21\times10^4$~cm$^{-3}$} and 5,000--13,000 K, respectively. 
The highest density of $2.1\times10^5$~cm$^{-3}$ occurs in the IRS 13 
region. Higher temperatures up to 13,000 K are found in the Bar 
region. Both the H30$\alpha$ and H92$\alpha$ line profiles are 
broadened due to  large velocity gradients along the line of sight 
and across the $2''$ beam produced by the dynamical motions 
near Sgr~A*. For the H92$\alpha$ line, the line width due to pressure 
broadening appears to be comparable to that of thermal Doppler 
broadening, while for the H30$\alpha$ lines, the pressure broadening 
is negligible.

Using the orbital parameters derived under the assumption that 
the three minispiral flows are in Keplerian motion, we calculated the 
3D geometry of the minispiral structure. We showed that  
the ionized flows of Northern Arm and Western Arc are almost 
coplanar, and the plane of the ionized flow in the Eastern Arm is 
nearly perpendicular to that of the Northern Arm and Western Arc. 
The Eastern Arm flow collides with the Northern Arm flow in the 
Bar region located 0.1--0.2 parsec south of and behind Sgr~A*. 
The high kinetic temperature gas in the Bar
is probably due to the heat in shocks generated by collisions.

We compared observed radial velocities in the Northern and Eastern Arms
with those computed from the Keplerian model with a mass of 
$4.2\times10^6~M_\odot$ of the SMBH at the position of Sgr~A*. We show 
that Keplerian motion dominates the kinematics of the minispiral arms. 
In some regions, significant deviations exist between the 
observed radial velocities and those computed from the Keplerian model. 
Near the IRS 9W and IRS 34 regions, the observed peculiar velocities in the 
minispiral are likely a result of the ionized gas being swept up by 
the strong stellar winds from nearby WR stars.

A total Lyman continuum flux of $3\times10^{50}$ phot.~s$^{-1}$ is 
needed to maintain the ionization of the gas in Sgr~A West with an 
inferred ionized mass of 350~$M_\odot$, which requires $\sim$250 
O9-type (ZAMS) stars. About 10\% of the total number of UV photons is 
required to externally maintain the ionization of the high-density 
gas in the Northern and Eastern Arms, which have an ionized mass of 
12~$M_\odot$, less than 5\% of the total ionized mass \citep{lisz03}.

\acknowledgments
We are grateful to Carolann Barrett for her careful editing of our
manuscript. We also thank Harvey Liszt for discussing aspects of 
his 2003 Keplerian model and for encouraging us to make these SMA 
observations. The research has made use of NASA's Astrophysics Data 
System. Support for DPM was provided by NASA through Hubble 
Fellowship grant HST-HF-51259.01 awarded by the Space Telescope 
Science Institute, which is operated by the Association of 
Universities for Research in Astronomy, Inc., for NASA, 
under contract NAS 5-26555.

\clearpage
\onecolumn
\include{figures}
\include{table1}

\include{table2}
\include{table3}
\end{document}

%% file: figures.tex

\begin{figure}[ht]
\figurenum{1}
\centering
\includegraphics[angle=-90, scale=0.8]{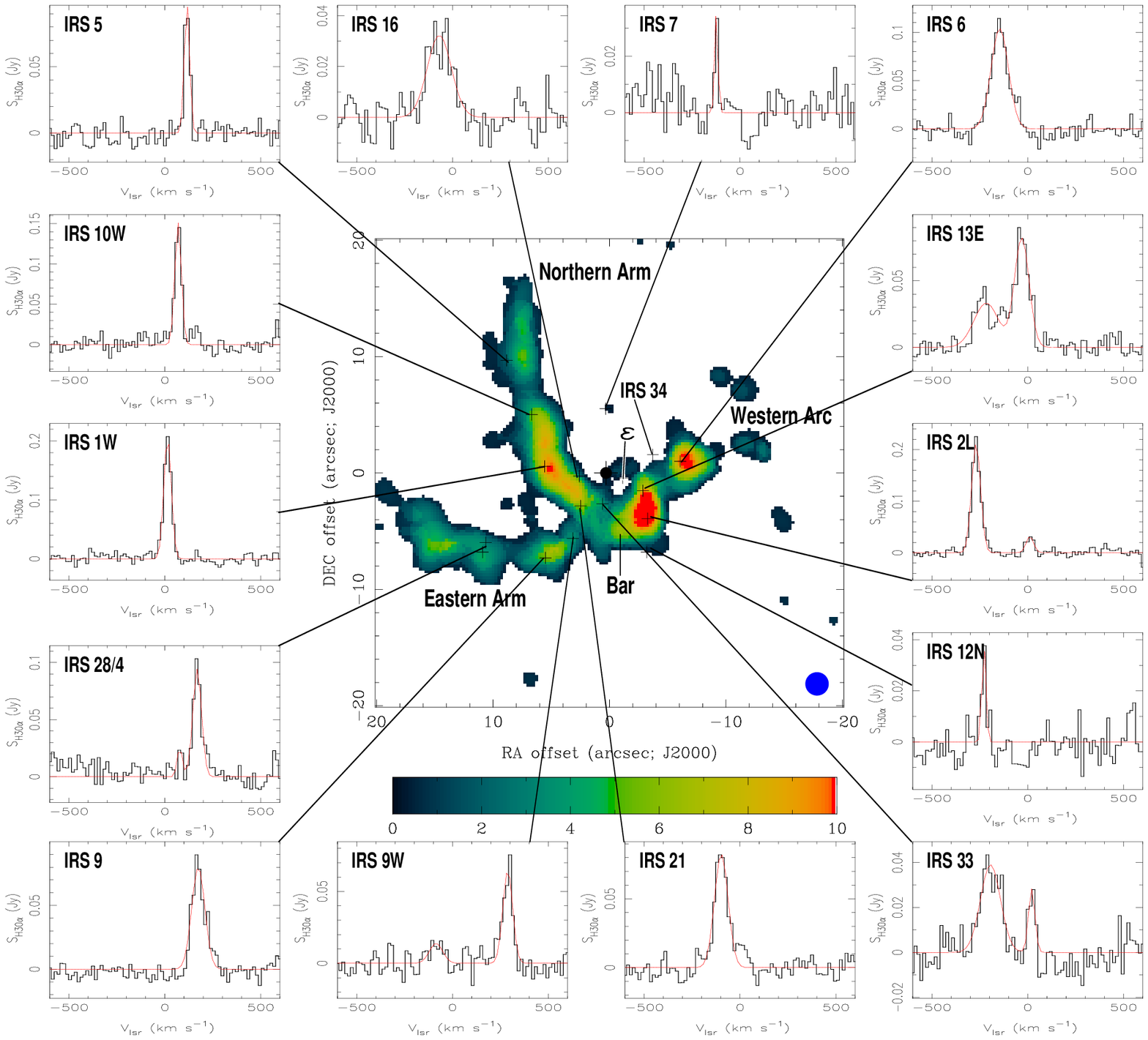}
\caption{\label{label1}
The image of the H30$\alpha$ line intensity integrated from 
the velocity range between $-360$ and $+345$ km~s$^{-1}$
with the cutoff of 30 mJy beam$^{-1}$ ($\sim4\sigma$) in
each of the channels. The 
color wedge shows the intensity in the range between 0 to 
10 Jy beam$^{-1}$ km s$^{-1}$. The beam size 
($\theta_{\rm FWHM}=2$\arcsec) is shown in the bottom right. 
The dot-cross at the coordinate origin marks 
the position of Sgr~A*, where ($\alpha_{J2000}=17^{\rm h}45^{\rm m}40.037^{\rm s}$, 
$\delta_{J2000}=-29^\circ00'28.11''$) defines the origin of the coordinate system. 
The plus signs mark the positions 
from which the H30$\alpha$ line spectra in the beam are 
taken and a few additional sources discussed in the text. The red curves are the Gaussian fits (multiple 
components in some cases) obtained from a least-squares 
analysis. The spectra are labeled with the names of 
their corresponding IRS sources. }
\end{figure}

\begin{figure}[ht]
\figurenum{2}
\centering
\includegraphics[angle=-90, scale=0.7]{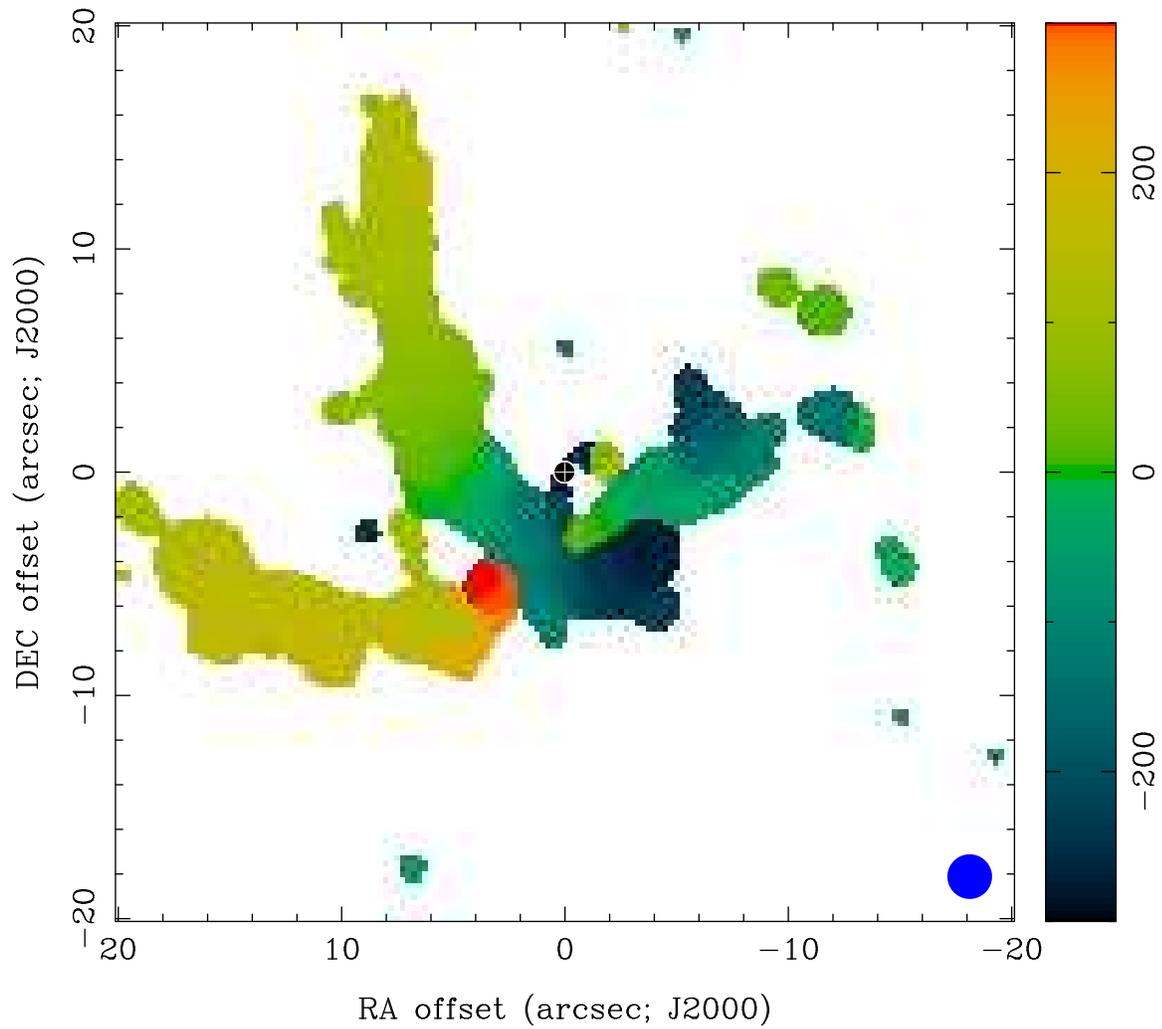}
\caption{\label{label2}
The distribution of the radial velocities determined from fitting 
the peak velocity of the H30$\alpha$ line image cube 
observed with the SMA in the central 40\arcsec~region. 
The color scale denotes velocity from --360 to +345 km s$^{-1}$.
The circular beam $\theta_{\rm FWHM}=2$\arcsec~is shown at the 
bottom right. The dot-cross at the coordinate origin 
marks the position of Sgr A*.}
\end{figure}

\begin{figure}[ht]
\figurenum{3}
\centering
\includegraphics[angle=0, scale=0.7]{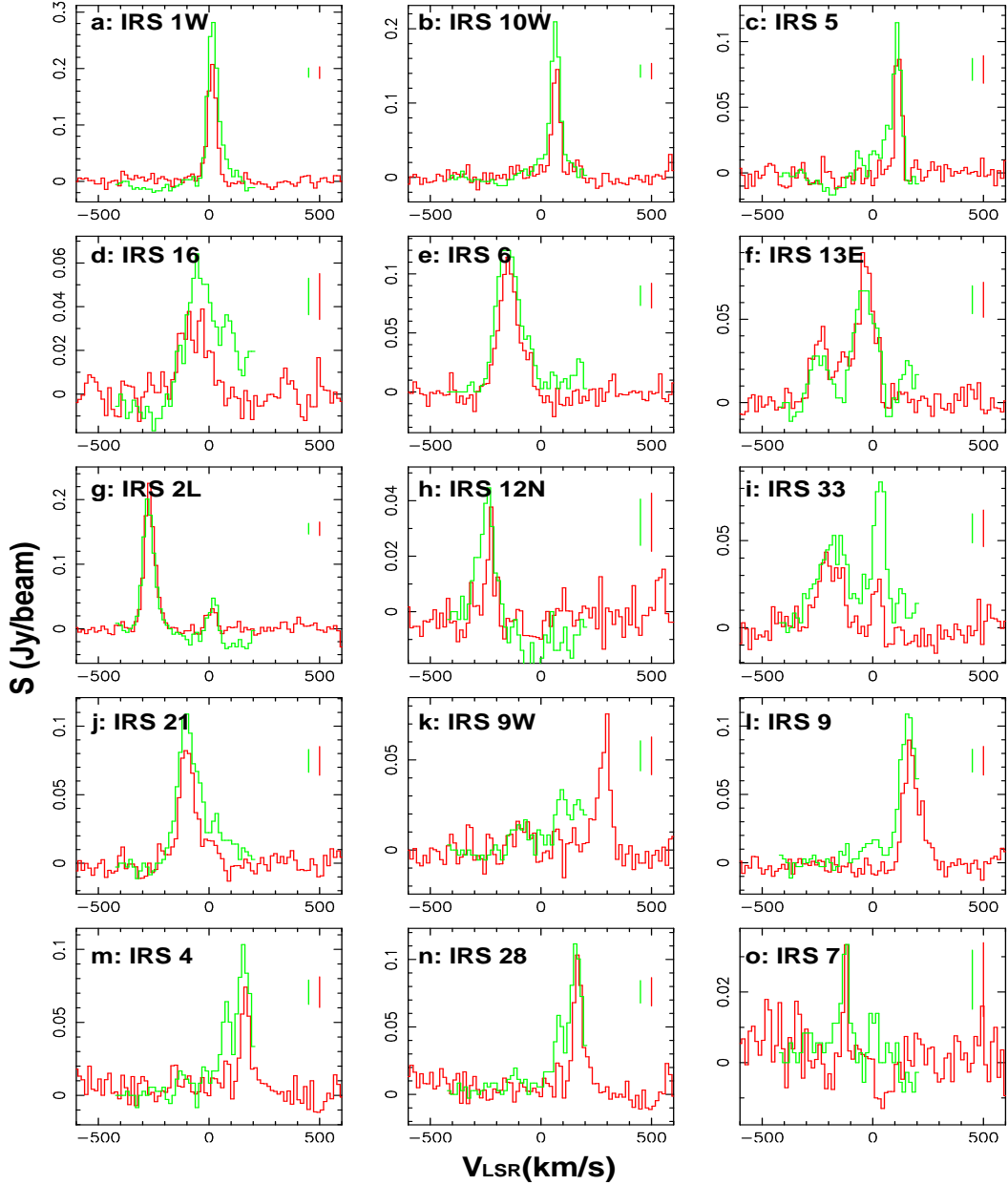}
\caption{\label{label3}
A comparison of the H30$\alpha$ line spectra (red) with the VLA 
H92$\alpha$ line spectra (green) taken from the selected IRS source 
regions with a resolution of $\theta_{\rm FWHM}=2$\arcsec. The VLA 
H92$\alpha$ spectra have been multipled by a factor of 
$\nu_{\rm H30\alpha}/\nu_{\rm H92\alpha}=28$. The vertical 
bars indicate the uncertainty of $3\sigma$. 
}
\end{figure}
\begin{figure}[ht]
\figurenum{4}
\centering
\includegraphics[angle=-90, scale=0.34]{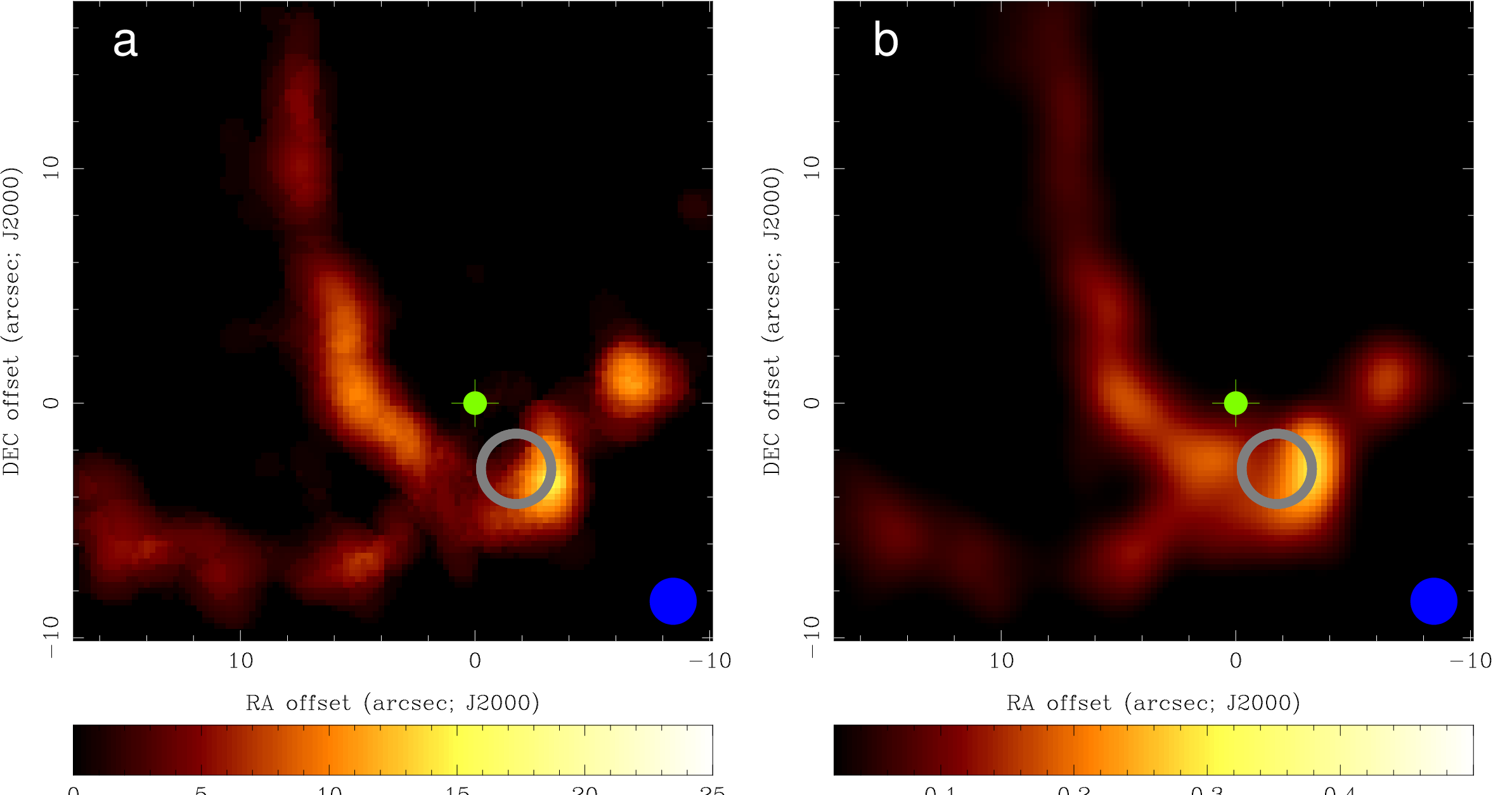}
\hfil
\centering
\includegraphics[angle=-90, scale=0.45]{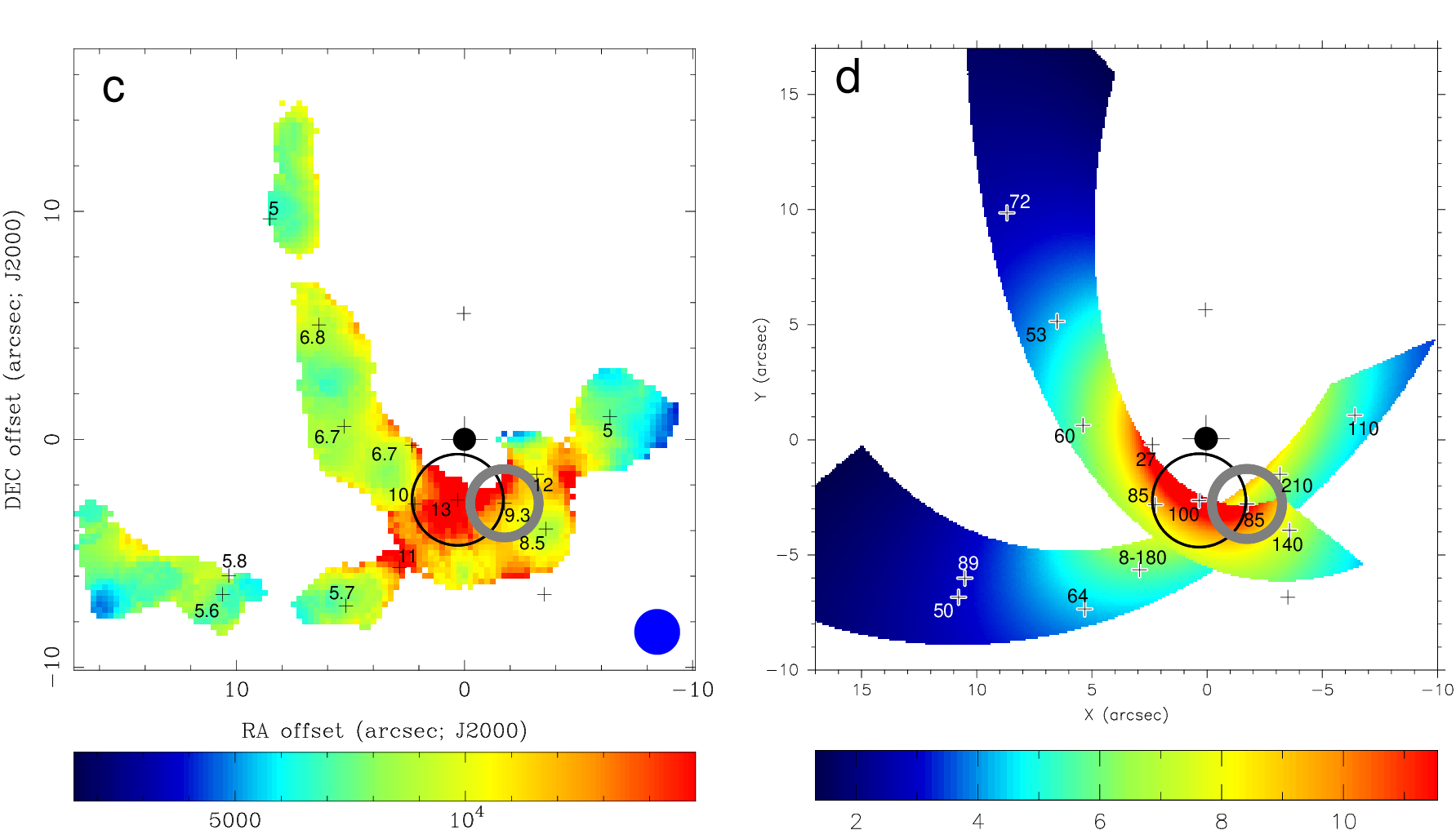}
\caption{\label{label4} {\bf a}) Integrated H30$\alpha$ line 
emission observed at 1.3 mm with the SMA. {\bf b}) Radio 
continuum image made from VLA observations at 1.3 cm with a 
resolution of $\theta_{\rm FWHM}=2$\arcsec. The radio emission 
associated with Sgr A* and the X-ray transient (J174540.0290031) 
has been removed, but the position of Sgr A* is marked by the 
dot-cross at the coordinate origin in each of the figures. 
{\bf c}) Pseudo-color shows the LTE temperature ($T_e^*$) derived 
from images of H30$\alpha$ line emission and radio continuum 
based on  Equation 1. {\bf d})~The image of the specific 
kinetic energy computed using the Keplerian orbital
model described in Table 5 of \cite{zhao09}. The color  
scale shows the specific kinetic energy in units of 
$10^{14}$ erg g$^{-1}$. In Figures 3c and 3d, 
the electron temperature ($T_e$ in 10$^3$ K) and the electron density 
($n_e$ in 10$^{3}$ cm$^{-3}$) are shown at the positions of selected 
IR sources. These numbers are derived using the isothermal
homogeneous HII model with the non-LTE effects. The results 
are tabulated in Table 3.  The thick gray circles in ({\bf a}),({\bf b}),
({\bf c}) and ({\bf d}) indicate the well-known minicavity 
region \citep{yusef89}; the black circles in ({\bf c}) and ({\bf d}) 
indicate high-temperature regions. }
\end{figure}

\begin{figure}[ht]
\figurenum{5}
\centering
\vfil 
\includegraphics[angle=0, scale=1.0]{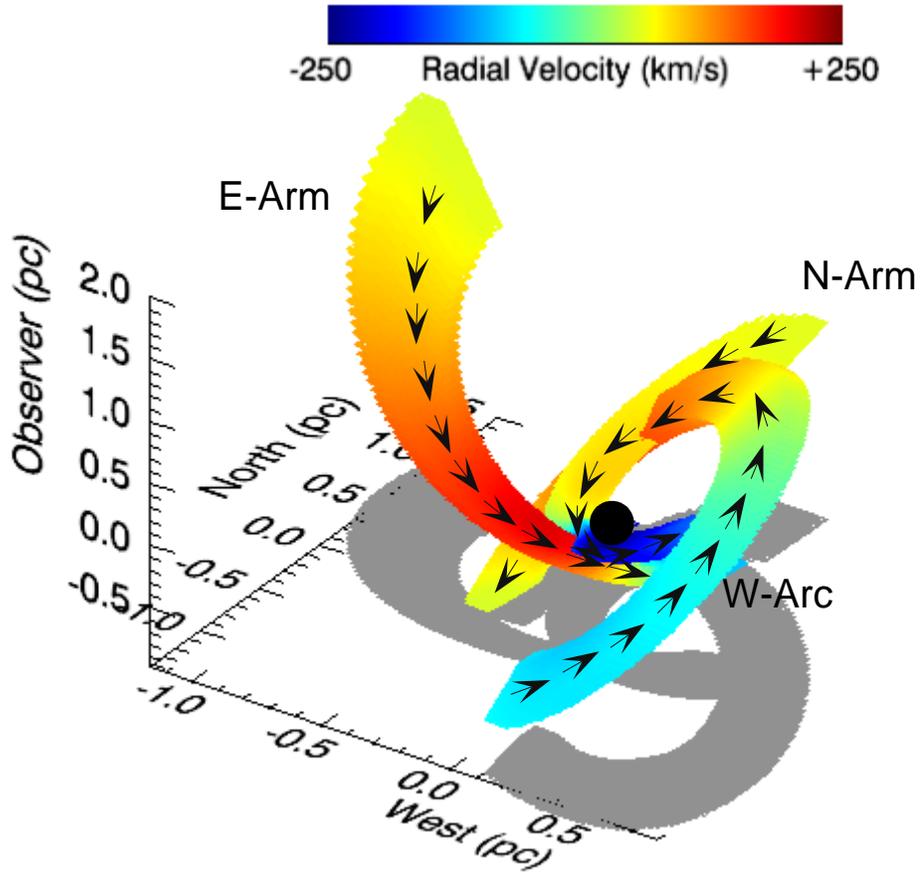}
\caption{3D view of the minispiral structure and velocity
along each of the minispiral arms in the nucleus of the Galaxy 
centered at Sgr A* (black dot), which are computed using the 
Keplerian model described in \cite{zhao09}. The orbital 
parameters were derived from fitting the VLA H92$\alpha$ line 
data and are summarized in Table 5 of \cite{zhao09}. 
The  arrows indicate the direction of the ionized flows 
along the orbits. The color illustrates the radial velocity between 
$-$250 to +250 km s$^{-1}$ (color wedge at top).
The gray shadow indicates the loci of the ionized flows 
projected on the sky plane.  }
\end{figure}

\begin{figure}[ht]
\figurenum{6}
\centering
\includegraphics[angle=0, scale=0.7]{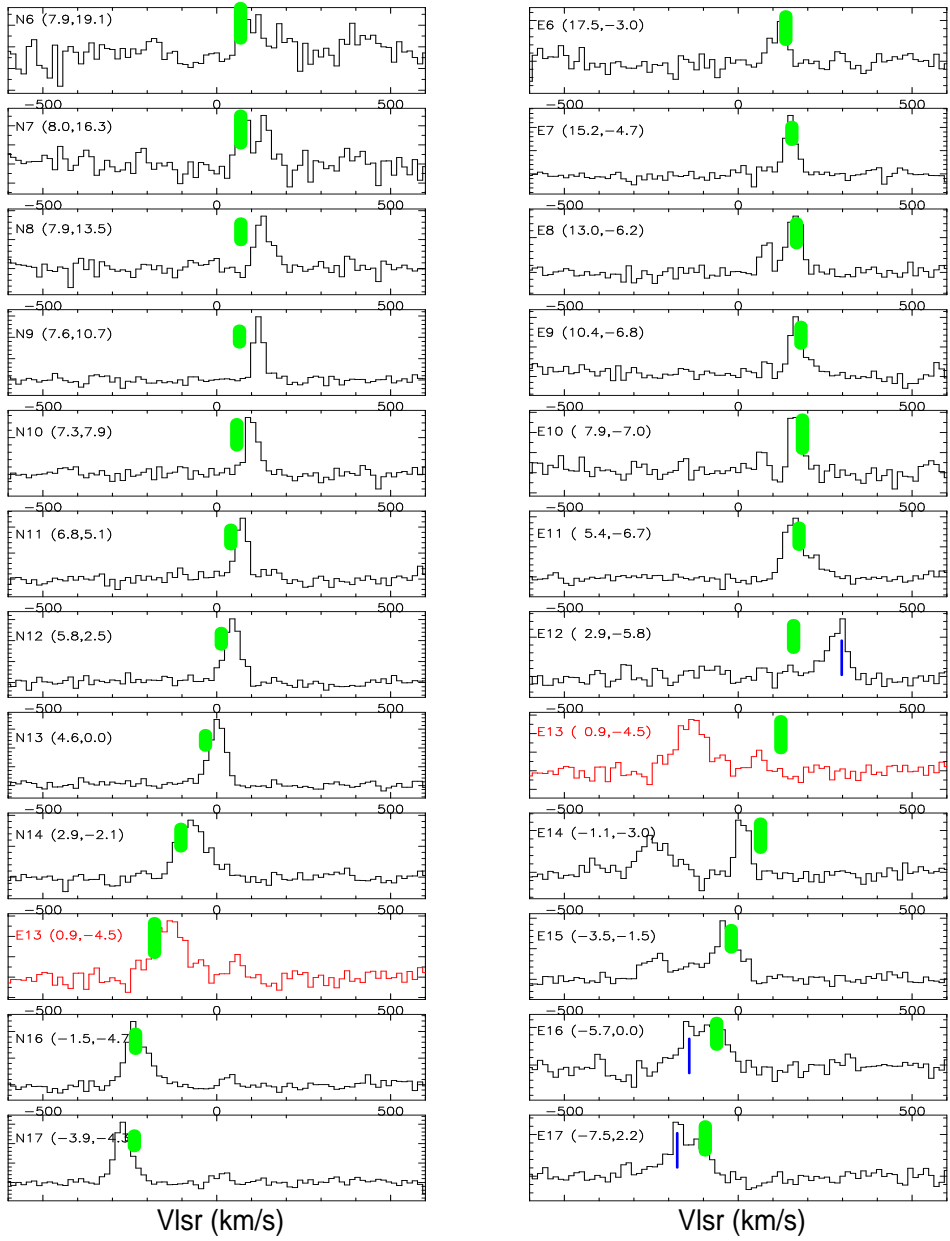}
\caption{\label{label6}
Left: Spectra of the H30$\alpha$ line taken along the Northern Arm.
Right: Spectra of the H30$\alpha$ line taken along the Eastern Arm.
The locations where the spectra were taken are shown in Figure 7. The green 
bars mark the radial velocities along the Northern and Eastern Arms 
computed using the Keplerian model with the orbital parameters 
given in Table 5 of \cite{zhao09}, assuming that the mass of the 
SMBH is $4.2\times10^6~M_\odot$ and its radial velocity $V_z=0$ km s$^{-1}$ 
with respect to the LSR. The thin blue bars (in panels E12, E16, and E17) 
indicate peculiar velocities possible due to interactions with the 
strong stellar winds from the nearby WR stars (see Section 3.2.3).
Note that the spectrum marked ``E13'' and shown in red in both
columns comes from a common position, with the appropriate Keplerian 
velocity marked for each arm.}
\end{figure}

\begin{figure}[ht]
\figurenum{7}
\centering
\includegraphics[angle=-90, scale=0.7]{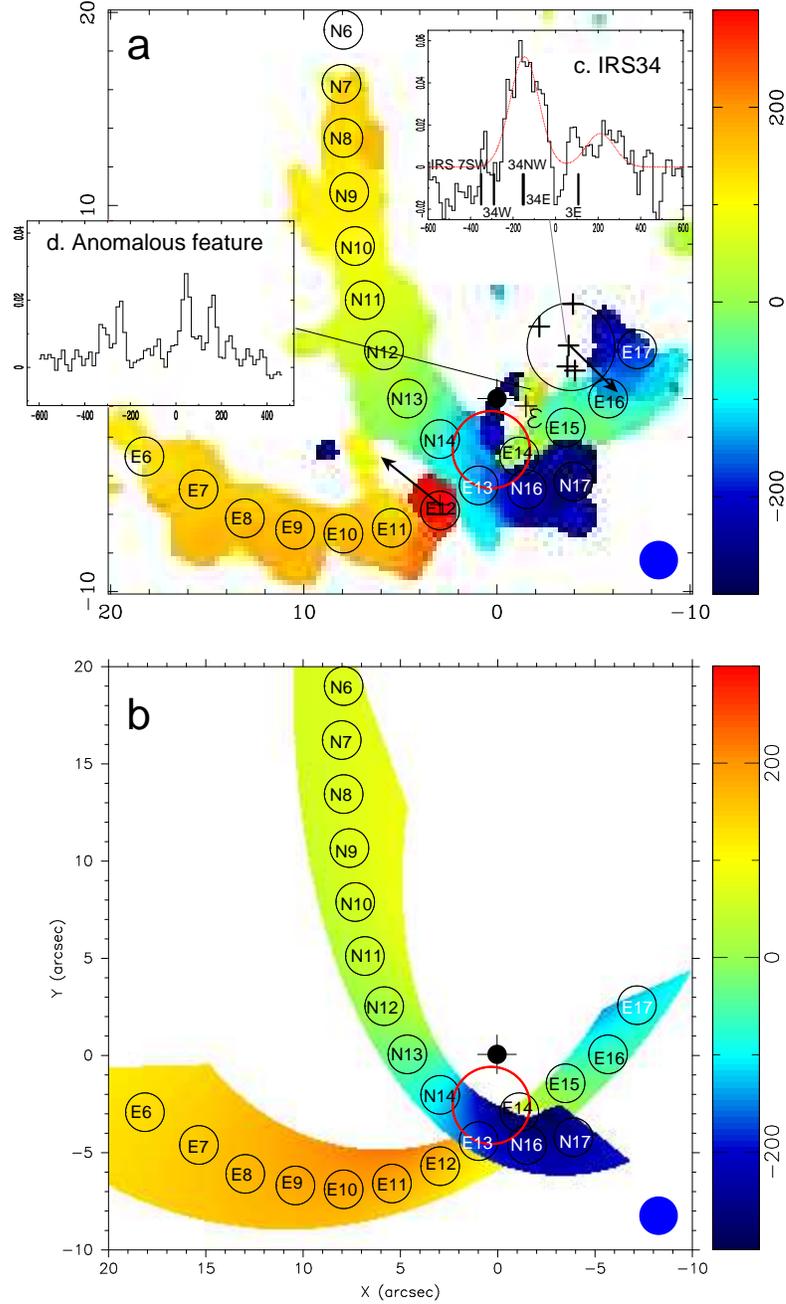}
\caption{\label{label7} 
A comparison between the observed radial velocities ({\bf a}) and 
those computed from the Keplerian model ({\bf b}). The black 
circles along the two spiral arms mark the locations of the 
H30$\alpha$ spectra displayed in Figure 6. The red circle indicates 
the high temperature region. The dot-cross marks the 
position of Sgr~A* at the coordinate origin. The inset ({\bf c}) 
is the line spectrum of H30$\alpha$ toward the IRS 34 region 
indicated with a large black circle, which includes the  five massive stars 
IRS 7SW (WN8), IRS 34W (Ofpe/WN9), IRS 34E (O9--9.5I), IRS 34NW (WN7), and 
IRS 3E (WC5/6) \citep{paum06}. The mean proper motion of the first 
four stars (and a fifth star without proper motion)
is indicated with an arrow near the E16 region in ({\bf a}). 
The radial velocities of the five stars are marked with bars in inset 
({\bf c}). The proper motion of the IRS 9W star (WN8) from 
\cite{paum06} is indicated with an arrow near the E12 region in 
({\bf a}). The inset ({\bf d}) shows the H30$\alpha$ line spectrum toward
an anomalous feature ($-1.5''$, $0.5''$) northwest of Sgr~A* and
 $1''$ north of 
$\varepsilon$.  
}
\end{figure}

\begin{figure}[ht]
\figurenum{8}
\centering
\includegraphics[angle=-90, scale=0.75]{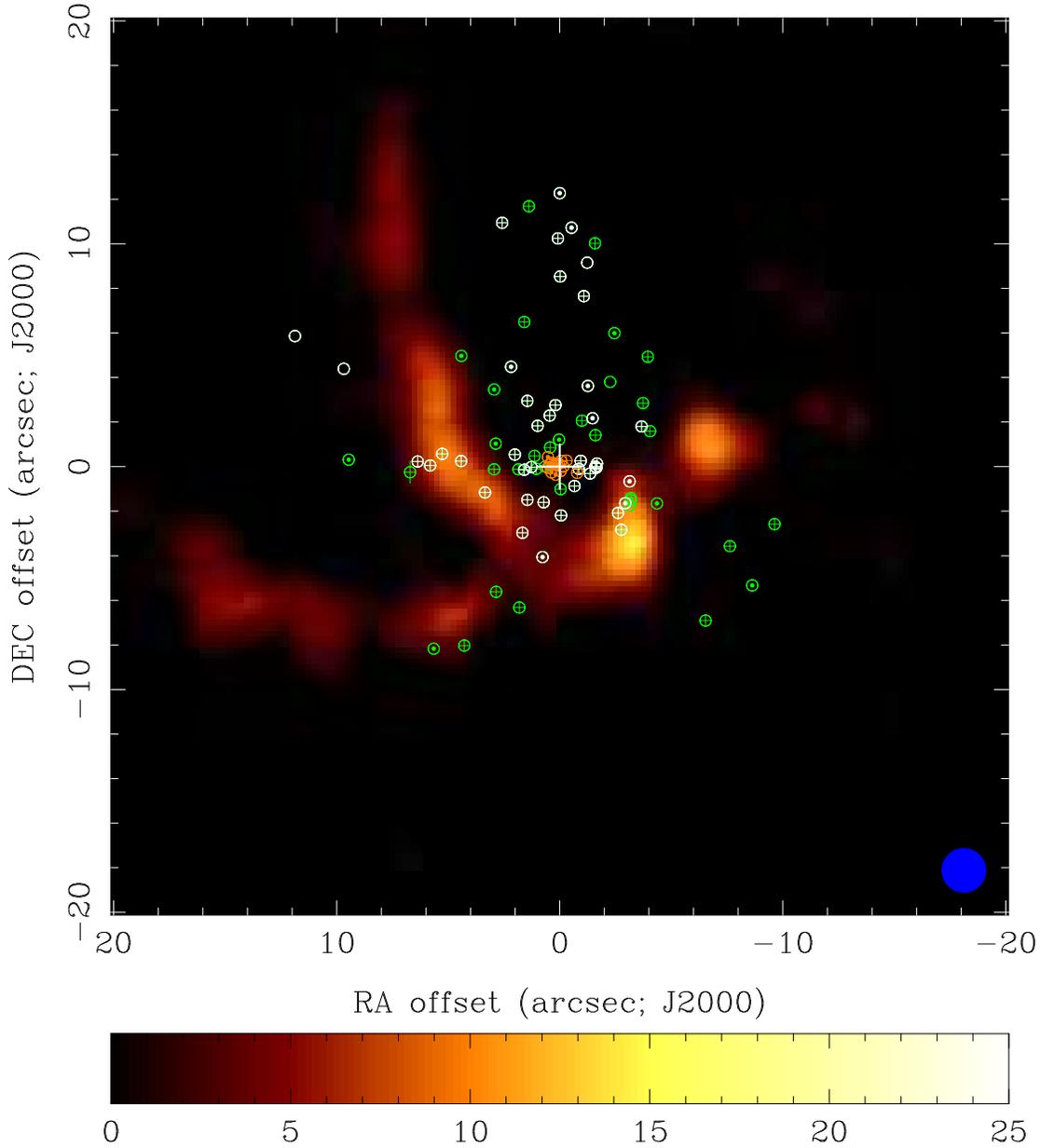}
\caption{\label{label8}
The 90 massive stars of \cite{paum06} are overlaid on the H30$\alpha$ 
line image. Among the 90 massive stars, 14 OB stars (orange),  
known as S stars, are located within a radius of $0.85''$ from 
Sgr A* (white plus sign),  41 OB stars (white) are outside the S cluster, 
and the rest are Ofpe/WN9 WR stars (green). The cross- and 
dot-circles represent the stars with proper motion data, and the open circles indicate stars
with no proper motion data. 
}
\end{figure}

%% file: table1.tex
\begin{deluxetable}{lcrcl}
\tablenum{1}
\tabletypesize{\scriptsize}
\tablewidth{0pt}
\tablecaption{Summary of the H30$\alpha$ Line Observations at $\nu=231.901$ GHz \label{tbl-1}}
\tablehead{
\colhead{Date}&
\colhead{$t_{obs}$ on Sgr A*}&
\colhead{$T_{sys}$}&
\colhead{Array\tablenotemark{{\rm a}}}&
\colhead{Calibrators}
\\
\colhead{}&
\colhead{(hr)}&
\colhead{(K)}&
\\
}
\startdata
2006-04-08 &  1.79 & 96  & C &3C273$^{\rm b}$, Callisto$^{\rm c}$ \\
2006-05-28 &  3.99 & 173 & E & 3C279$^{\rm b}$, Uranus$^{\rm b}$,
                                    J1924-292$^{\rm b}$, Callisto$^{\rm b,c}$ \\
2006-07-17 &  2.19 & 146 & V & 3C279$^{\rm b}$, J1924-292$^{\rm b}$,
                                    Callisto$^{\rm c}$ \\
2007-03-31 &  3.67 & 108 & C &3C273$^{\rm b}$, Callisto$^{\rm c}$ \\
2007-04-01 &  3.35 & 126 & C &3C273$^{\rm b}$, Callisto$^{\rm c}$ \\
2007-04-03 &  1.99 &  93 & C &3C273$^{\rm b}$, Callisto$^{\rm c}$ \\
2007-04-04 &  1.93 & 103 & C &3C273$^{\rm b}$, Callisto$^{\rm c}$ \\
2007-08-14 &  3.63 & 115 & E &3C279$^{\rm b}$, NRAO 530$^{\rm c}$ \\
2007-08-20 &  4.94 &  95 & E &3C454.3$^{\rm b}$, NRAO 530$^{\rm c}$ \\
2007-08-31 &  4.19 & 134 & E &3C454.3$^{\rm b}$, NRAO 530$^{\rm c}$ \\
2008-05-05 &  2.67 & 156 & C &3C273$^{\rm b}$, Ganymede$^{\rm c}$ \\
\enddata
\tablenotetext{{\rm a}} {Array configuration: C = Compact; 
E = Extended; V = Very Extended}
\tablenotetext{{\rm b}} {Calibrator: bandpass}
\tablenotetext{{\rm c}} {Calibrator: flux density scale}
\end{deluxetable}

%% file: table2.tex
\begin{deluxetable}{lrrrrrrcrr}
\tablenum{2}
\tabletypesize{\scriptsize}
\tablecaption{The H30$\alpha$ \& H92$\alpha$ Line Measurements} 
\tablewidth{0pt}
\tablehead{
\multicolumn{9}{c}{Radio Properties in IRS Regions} \\
\cline{4-6} \\
\colhead{Region}&
\colhead{$ {\Delta\alpha} $} &
\colhead{$ {\Delta\delta} $ } &
\colhead{$ {S_{\rm H30\alpha}}$}&
\colhead{$ {S_{\rm 22GHz}}$}&
\colhead{$\displaystyle {S_{\rm H30\alpha}\over S_{\rm 22GHz}}$}&
\colhead{$\displaystyle {S_{\rm H92\alpha}\over S_{\rm 22GHz}}$}&
\colhead{$\displaystyle {S_{\rm H30\alpha}\over S_{\rm H92\alpha}}$}&
\colhead{$V_{\rm LSR}$\tablenotemark{a}}&
\colhead{$\Delta V_{\rm FWHM}$\tablenotemark{a}}
\\
\colhead{}&
\colhead{(arcsec)}&
\colhead{(arcsec)}&
\colhead{(Jy beam$^{-1}$)}&
\colhead{(Jy beam$^{-1}$)}&
\colhead{}&
\colhead{}&
\colhead{}&
\colhead{(km s$^{-1}$)}&
\colhead{(km s$^{-1}$)}
\\ 
}
\startdata
IRS 1W&5.27&0.57& 0.21$\pm$0.008  &0.16$\pm$0.008 &1.3$\pm$0.08&0.06$\pm$0.005 &21$\pm$1&  14$\pm$1&46$\pm$2\\
IRS 10W&6.38 &5.02 & 0.15$\pm$0.018 &0.10$\pm$0.003&1.5$\pm$0.19&0.07$\pm$0.004 &21$\pm$1& 69$\pm$2&39$\pm$5\\
IRS 5&8.53 &9.63 & 0.10$\pm$0.019   &0.04$\pm$0.006 &2.5$\pm$0.6&0.1$\pm$0.02 &25$\pm$2 &115$\pm$3&33$\pm$8\\
IRS 16&2.30 &--0.26 &0.03$\pm$0.005 &0.08$\pm$0.015 &0.4$\pm$0.1&0.026$\pm$0.008 &15$\pm$4& --69$\pm$10&141$\pm$27\\
IRS 6&--6.37 &1.00 & 0.12$\pm$0.011 &0.15$\pm$0.006&0.80$\pm$0.08&0.034$\pm$0.003&23$\pm$2&--146$\pm$5&109$\pm$13\\
IRS13E&--3.17&--1.53&0.08$\pm$0.006&0.25$\pm$0.030&0.32$\pm$0.05&0.010$\pm$0.002&32$\pm$4&--31$\pm$3&86$\pm$8\\
IRS 2L&--3.57&--3.93& 0.21$\pm$0.011&0.23$\pm$0.030&0.91$\pm$0.13&0.031$\pm$0.005&29$\pm$1&--269$\pm$1&55$\pm$3\\
IRS 12N&--3.50&--6.80&0.04$\pm$0.010&0.04$\pm$0.002&0.76$\pm$0.22&0.036$\pm$0.013&21$\pm$6&--229$\pm$7&40$\pm$17\\
IRS 33&0.30&--2.65& 0.04$\pm$0.008&0.19$\pm$0.024  &0.21$\pm$0.05&0.009$\pm$0.003&24$\pm$5&--195$\pm$12&126$\pm$29\\
IRS 21&2.18 &--2.83 & 0.08$\pm$0.009&0.16$\pm$0.018&0.50$\pm$0.08&0.023$\pm$0.005&22$\pm$3&--99$\pm$4&80$\pm$11\\
IRS 9W&2.85 &--5.62 & 0.06$\pm$0.007&0.11$\pm$0.011&0.57$\pm$0.09&\nodata\tablenotemark{b}&\nodata\tablenotemark{b}&288$\pm$4&60$\pm$8\\
IRS 9&5.20 &--7.30  & 0.08$\pm$0.009&0.09$\pm$0.009&0.85$\pm$0.13&0.043$\pm$0.008&20$\pm$2&174$\pm$5&84$\pm$11\\
IRS 4& 10.33&--5.98  & 0.07$\pm$0.012&0.05$\pm$0.006&1.4$\pm$0.3&0.07$\pm$0.02&20$\pm$3&164$\pm$4&47$\pm$10\\ 
IRS 28& 10.60 &--6.80&0.10$\pm$0.010&0.07$\pm$0.006&1.5$\pm$0.2&0.06$\pm$0.01&25$\pm$2&168$\pm$4&49$\pm$22\\
IRS 7 &0.03 &5.52 & 0.03$\pm$0.010&0.15$\pm$0.001&2.0$\pm$0.6 &0.07$\pm$0.03 &28$\pm$8&--124$\pm$10&29$\pm$22\\
Minicavity\tablenotemark{g}&--1.75&--2.80& 0.08$\pm$0.006&0.18$\pm$0.020&0.45$\pm$0.07&0.016$\pm$0.002&29$\pm$2&6$\pm$2&41$\pm$4\\
\\
\tableline
\\
\multicolumn{9}{c}{Radio Properties of Sgr A West} \\
\cline{4-6} \\
\colhead{Component} &  
\multicolumn{2}{c}{$\int S_{\rm H30\alpha} dV$}&
\multicolumn{2}{c}{$\int S_{\rm H92\alpha} dV$}&
\multicolumn{2}{c}{$ S_{\rm 22GHz} $}& 
\multicolumn{2}{c}{$ Area$} \\
\colhead{}&  
\multicolumn{2}{c}{($\rm Jy$ \kms)}&
\multicolumn{2}{c}{($\rm Jy$ \kms)}& 
\multicolumn{2}{c}{($\rm Jy$)}& 
\colhead{(arcmin$^2$)}&
\colhead{(pc$^2$)}
\\
\\
\tableline
\\
Sgr A West total\tablenotemark{c}& 
\multicolumn{2}{c}{1700$\pm$300\tablenotemark{d}} &
\multicolumn{2}{c}{60$\pm$7} &
\multicolumn{2}{c}{16.3$\pm$0.2}& 
1.0$\pm$0.1&
5.5$\pm$0.5   \\
A\tablenotemark{e}  &
\multicolumn{2}{c}{300$\pm$20}&
\multicolumn{2}{c}{19$\pm$2}  &
\multicolumn{2}{c}{7.1$\pm$0.1} &
0.10$\pm$0.01&
0.55$\pm$0.05 \\
B\tablenotemark{f}  &
\multicolumn{2}{c}{1400$\pm$300}&
\multicolumn{2}{c}{41$\pm$7}&
\multicolumn{2}{c}{9.2$\pm$0.2} &
1.0$\pm$0.1&
5.5$\pm$0.5   \\
\enddata
\tablenotetext{a}{Determined from the H30$\alpha$ line.}
\tablenotetext{b} {No H92$\alpha$ data at the line velocity.}
\tablenotetext{c}{The overall Sgr A West region (80\arcsec$\times$45\arcsec).}
\tablenotetext{d}{The line flux determined from the IRAM 30m observation.}
\tablenotetext{e}{The bright H30$\alpha$ line emission region 
(the Northern and Eastern Arms in Fig. 2).}
\tablenotetext{f}{The residual of Sgr A West after taking out the contribution 
from component A.}
\tablenotetext{g}{The line and continuum intensities listed in this row
reflect averaged values of the emission from both the rim and the void of
the minicavity in the  2\arcsec\/ beam that is similar to
the diameter of the minicavity.} 
\end{deluxetable}

%% file: table3.tex
\begin{deluxetable}{lccccccc}
\tablenum{3}
\tabletypesize{\scriptsize}
\tablecaption{Physical Conditions of Ionized Gas in IRS Regions} 
\tablewidth{0pt}
\tablehead{
\colhead{Region} &
\colhead{$ {T_{\rm e}} $} &
\colhead{$ {n_{\rm e}} $} &
\colhead{$ {EM}$}&
\colhead{$ {\rm \tau_C({\rm 22GHz})}$}&
\colhead{${\displaystyle \left(T_{\rm e}\over T_{\rm e}^*\right)^{\tablenotemark{\dagger}}}$}&
\colhead{$\displaystyle \left(\Delta V_{t}\over             
\Delta V_{\rm FWHM}\right)$} &
\colhead{$\displaystyle \left( \Delta V_{\rm P}^{\rm H92\alpha}
\over\Delta V_{\rm D}^{\rm TH}
\right)^{\tablenotemark{\ddagger}}$} 
\\
\colhead{}&
\colhead{($10^3$ K)}&
\colhead{($10^4$ cm$^{-3}$)}&
\colhead{($10^7$ cm$^{-6}$ pc)}&
\colhead{($10^{-3}$)}&
\colhead{}&
\colhead{}&
\colhead{}
\\ 
}
\startdata
IRS 1W   &6.7$\pm$0.5&6.0$\pm$0.4&1.8$\pm$0.1&14.$\pm$1.1&0.71&0.88& 0.82\\
IRS 10W  &6.8$\pm$0.6&5.3$\pm$0.4&1.1$\pm$0.1&8.5$\pm$1.0&0.71&0.85& 0.72\\
IRS 5    &5.0$\pm$1.1&7.2$\pm$1.2&0.4$\pm$0.1&4.7$\pm$1.3&0.68&0.78& 1.2\\
IRS 16   &6.7$\pm$2.0&2.7$\pm$0.7&0.8$\pm$0.2&6.4$\pm$2.7&0.68&0.99& 0.37\\
IRS 6    &5.0$\pm$0.6&11.$\pm$1.1&1.6$\pm$0.1&18.$\pm$2.3&0.71&0.96& 1.8\\
IRS 13E  &12.$\pm$2.5&21.$\pm$3.6&3.2$\pm$0.5&12.$\pm$1.9&0.84&0.84& 2.0\\
IRS 2L   &8.5$\pm$1.3&14.$\pm$1.4&2.7$\pm$0.4&16.$\pm$2.9&0.78&0.80& 1.7\\
IRS 12N  &12.$\pm$4.7&2.8$\pm$1.0&0.5$\pm$0.2&1.8$\pm$0.6&0.77&0.80& 0.27\\
IRS 33   &13.$\pm$3.3&10.$\pm$3.0&2.6$\pm$0.3&8.5$\pm$1.8&0.81&0.97& 0.92\\
IRS 21   &10.$\pm$1.8&8.5$\pm$1.3&2.0$\pm$0.3&9.1$\pm$2.0&0.78&0.94& 0.92\\
IRS 9W&11.$-$12.  &0.8$-$18   &1.3$-$1.3  &5.5$-$4.9&0.74$-$0.83&0.76$-$0.93&1.7$-$0.08\\
IRS 9    &5.7$\pm$0.9&6.4$\pm$0.8&1.0$\pm$0.2&9.7$\pm$1.8&0.69&0.97& 0.97\\
IRS 4    &5.8$\pm$1.3&5.0$\pm$1.0&0.5$\pm$0.1&4.8$\pm$1.4&0.68&0.91& 0.75\\
IRS 28   &5.6$\pm$0.8&8.9$\pm$1.0&0.7$\pm$0.1&6.8$\pm$1.1&0.71&0.86& 1.4\\
IRS 7    &7.0$\pm$2.5&8.8$\pm$2.7&0.2$\pm$0.1&1.2$\pm$0.5&0.74&0.64& 1.2\\
Minicavity&9.3$\pm$0.9&8.5$\pm$0.7&2.0$\pm$0.3&10.$\pm$2.0&0.77&0.78& 0.96\\
\enddata
\tablenotetext{\dagger} {$\displaystyle {T_{\rm e}\over T_{\rm e}^*}
\approx\left[b_{30}(1-{1\over2}\beta_{30}\tau_{\rm C}(232{\rm ~GHz})\right]^{0.87}
\approx b_{30}^{0.87}$ for a very small value of $\tau_{\rm C}(232{\rm ~GHz})$ in a region
at the Galactic center.}
\tablenotetext{\ddagger} {$\Delta V_{\rm D}^{\rm TH} = \displaystyle \sqrt{8 {\rm ln} (2) k T_e\over M_H}$ is the thermal Doppler line width.} 
\end{deluxetable}